\documentclass{article}
\usepackage{setspace}
\usepackage[utf8]{inputenc}
\usepackage{amsfonts,amsmath,amssymb,mathtools}
\usepackage[left=2cm,right=2cm,top=2cm,bottom=2cm,bindingoffset=0cm]{geometry}
\usepackage{graphicx}
\usepackage{authblk}
\usepackage{xcolor}
\usepackage{hyperref}
\usepackage{authblk}
\usepackage{slashed}
\usepackage{braket}
\usepackage{comment}
\usepackage{pb-diagram}
\usepackage{wasysym}
\usepackage{amsthm}
\usepackage{cancel}
\usepackage{tikz}\usetikzlibrary{tikzmark}
\usetikzlibrary{fadings} 
\usetikzlibrary{positioning}
\usetikzlibrary{backgrounds} 
\usetikzlibrary[backgrounds]
\usetikzlibrary{decorations.pathreplacing}
\usetikzlibrary{arrows}
\definecolor{airforceblue}{rgb}{0.36, 0.54, 0.66}	
\definecolor{beige}{rgb}{0.96, 0.96, 0.86}
\definecolor{bittersweet}{rgb}{1.0, 0.44, 0.37}
\definecolor{melon}{rgb}{0.99, 0.74, 0.71}
\definecolor{mustard}{rgb}{1.0, 0.86, 0.35}
\definecolor{lava}{rgb}{0.81, 0.06, 0.13}
\definecolor{magnolia}{rgb}{0.97, 0.96, 1.0}
\definecolor{lavendermist}{rgb}{0.9, 0.9, 0.98}
\definecolor{lavendergray}{rgb}{0.77, 0.76, 0.82}
\definecolor{palepink}{rgb}{0.98, 0.85, 0.87}
\definecolor{palesilver}{rgb}{0.79, 0.75, 0.73}
\definecolor{cadetgrey}{rgb}{0.57, 0.64, 0.69}
\definecolor{anti-flashwhite}{rgb}{0.95, 0.95, 0.96}
\colorlet{Light0anti-flashwhite}{anti-flashwhite!70!white}
\colorlet{Lightanti-flashwhite}{anti-flashwhite!50!white}
\colorlet{Light2anti-flashwhite}{anti-flashwhite!30!white}
\definecolor{linkcolor}{rgb}{0,0,1}
\definecolor{urlcolor}{rgb}{0,0,1}

\usepackage{tikz-cd}
\usetikzlibrary{decorations.markings}
\usetikzlibrary{positioning}
\usepackage{braids}
\usepackage{array}
\usepackage{ytableau}
\usepackage{longtable}
\usepackage{empheq}
\usepackage{multicol}
\usepackage{mathrsfs}
\usepackage{diagbox}
\usepackage[vcentermath]{youngtab}
\usepackage[natbib=true, backend = bibtex, style=numeric-comp, sorting=none]{biblatex}
\hypersetup{pdfstartview=FitH,  linkcolor=linkcolor,urlcolor=urlcolor,citecolor=urlcolor, colorlinks=true}

\newcommand\bem{\begin{pmatrix}}
\newcommand\eem{\end{pmatrix}}
\newcommand\beq{\begin{equation}}
\newcommand\eeq{\end{equation}}
\newcommand\beqs{\begin{equation*}}
\newcommand\eeqs{\end{equation*}}

\newcommand{\tr}{\text{tr}}

\date{}
\begin{document}

\title{\bf Chern-Simons perturbative series revisited}
\author[1,2]{{\bf E.~Lanina}\thanks{\href{mailto:lanina.en@phystech.edu}{lanina.en@phystech.edu}}}
\author[1,2,3]{{\bf A.~Sleptsov}\thanks{\href{mailto:sleptsov@itep.ru}{sleptsov@itep.ru}}}
\author[1,2]{{\bf N.~Tselousov}\thanks{\href{mailto:tselousov.ns@phystech.edu}{tselousov.ns@phystech.edu}}}

\vspace{5cm}

\affil[1]{Moscow Institute of Physics and Technology, 141700, Dolgoprudny, Russia}
\affil[2]{Institute for Theoretical and Experimental Physics, 117218, Moscow, Russia}
\affil[3]{Institute for Information Transmission Problems, 127994, Moscow, Russia}
\renewcommand\Affilfont{\itshape\small}
\maketitle

\vspace{-7cm}

\begin{center}
	\hfill ITEP-TH-14/21\\
	\hfill IITP-TH-11/21\\
	\hfill MIPT-TH-10/21
\end{center}

\vspace{5cm}

\begin{abstract}
A group-theoretical structure in a perturbative expansion of the Wilson loops in the 3d Chern-Simons theory with $SU(N)$ gauge group is studied in symmetric approach. A special basis in the center of the universal enveloping algebra $ZU(\mathfrak{sl}_N)$ is introduced. This basis allows one to present group factors in an arbitrary irreducible finite-dimensional representation. Developed methods have wide applications, the most straightforward and evident ones are mentioned. Namely, Vassiliev invariants of higher orders are computed, a conjecture about existence of new symmetries of the colored HOMFLY polynomials is stated, and the recently discovered tug-the-hook symmetry of the colored HOMFLY polynomial is proved.
\end{abstract}

\setcounter{equation}{0}
\section{Introduction}\label{intro}
This paper revisits perturbative studies of correlators in the Chern-Simons theory~\cite{Alvarez1994tt, Alvarez_1997, Alvarez_1993, Labastida1997uw, Guadagnini:1989kr, GUADAGNINI1990575}. The theory deserves attention for several reasons. It is particularly interesting for us due to its gauge invariant observables -- Wilson loop operators, which turn out to be connected with the colored HOMFLY polynomials of knots. These operators are long known to govern one of the most intriguing physical phenomena, such that the quark confinement in QCD \cite{Polyakov1976fu}. The Chern-Simons theory may play a role of an intermediate step to the complete understanding of physics of the Wilson loop observables in quantum field theories.

It is already much known about the Chern-Simons theory due to its connections with various topics of modern theoretical and mathematical physics: quantum field theory \cite{Witten1988hf,Chern1974ft,Guadagnini:1989kr,GUADAGNINI1990575,Kaul1991np,RamaDevi1992np,Ramadevi1993np,Ramadevi1994zb}, quantum qroups \cite{TURAEV1992865,Mironov2011aa,Anokhina2013ica,Anokhina2012ae}, 2d conformal WZW theories \cite{Kaul1991np,RamaDevi1992np,Ramadevi1993np,Zodinmawia2011oya,Zodinmawia2012kn}, topological strings \cite{Ooguri1999bv, Labastida2000zp, Labastida2000yw, Labastida2001ts, Marino2001re} and knot theory \cite{Alexander, Conway, Jones,Bishler2020fcy,Bishler2020wbq, Morton}. Here we utilize its connection with the theory of quantum knot invariants.

In this paper we study group-theoretical constituents of a perturbative expansion of the Wilson loops -- group factors; their definition is provided in Section~\ref{colored HOMFLY}. Currently, an explicit description of group factors is known up to the 6-th order for $\mathfrak{sl}_N$ fundamental and symmetric $[2]$ representations \cite{Sleptsov_2016}, and for any $\mathfrak{sl}_2$ representation \cite{chmutov_duzhin_mostovoy_2012}. The computational complexity increases drastically as the representation becomes bigger, which restricts the number of answers to be obtained. This makes the internal structure of the colored HOMFLY polynomials concealed for the present moment. Here \textit{we present a concrete algorithm} for calculating the HOMFLY group factors \textit{up to any order} and in \textit{any irreducible finite-dimensional representation}.

Group factors are not specific for the Chern-Simons theory, they arise in any non-Abelian gauge theory (see \cite{Groupfactors} as an example). Group factors are natural components of perturbative calculations, see for example calculation of higher loop corrections to the QCD beta-function \cite{4loopQCD, 5loopQCD}, renormalization theory in the Chern-Simons model \cite{GUADAGNINI2016238}, correlators in BF model \cite{GUADAGNINI2020114987} and recent papers on $\mathcal{N} = 4$ SYM \cite{SYM, SYMgroupfactors}.
\\
\\
This paper is organized as follows. In Section \ref{colored HOMFLY} we define the \textit{basic notions} of the colored HOMFLY polynomial, its group factors and Vassiliev invariants. Subsection~\ref{embed into gl} is devoted to the \textit{main tools} that we use in the study: the $\mathfrak{sl}_N$ Casimir eigenvalues and their embedding to the space of $\mathfrak{gl}_{\infty}$ Casimir eigenvalues. In Subsection \ref{HOMFLYsym} we discuss the symmetries of the colored HOMFLY polynomial that we use to restrict the form of the group factors. In Subsection~\ref{ConstrWeightSystem} we present an explicit construction of the HOMFLY group factors by the use of a special basis in the center of the universal enveloping algebra $ZU(\mathfrak{sl}_N)$. We provide group factors up to the 9-th level explicitly. Section~\ref{Implications} is devoted to some straightforward implications of the obtained HOMFLY group structure.

\setcounter{equation}{0}
\section{Colored HOMFLY polynomial}\label{colored HOMFLY}
Colored HOMFLY polynomials are topological invariants of knots and links. They generalize many known polynomial knot invariants, such as the well-known Jones and Alexander polynomials, quantum Reshetikhin-Turaev invariants of $\mathfrak{sl}_N$. An essence of this generalization is a clever analytical continuation by introducing a new variable $a = q^N$, which allows one to connect the colored HOMFLY polynomials with quantum knot invariants not only of Lie groups, but also of Lie supergroups \cite{queffelec2015note}.

The colored HOMFLY polynomial can be introduced in several self-consistent ways. Precise \textit{mathematical definitions} of the HOMFLY polynomial can be given in two ways. \\
$\bullet$ By the notion of \textit{the Kontsevich integral}, which one can see in Ch.8 of~\cite{chmutov_duzhin_mostovoy_2012}. It has a straightforward connection with a perturbative expansion of the HOMFLY polynomial, which is shown in more details below. \\
$\bullet$ With the use of \textit{the Reshetikhin-Turaev approach}, which is well-illustrated in~\cite{dhara2019multi}.  \\
For theoretical physicists it is more convenient to use some sort of \textit{physical definitions} of the colored HOMFLY polynomial. They followed from the classical Witten's work~\cite{Witten1988hf}. \\
$\bullet$ The first definition can be introduced by the use of \textit{the 2d conformal WZW theories}. This connection is well-established, for example, in~\cite{anokhina2015rmatrix}. \\
$\bullet$ As it has already been mentioned in Introduction, the second one can be stated through the connection between the colored HOMFLY polynomial and \textit{the Chern-Simons theory}. To simplify the subsequent description of obtained results, we utilize exactly this one. \\ \\
It was a breakthrough when the connection between knots and topological quantum field theories was established \cite{Witten1988hf}. In particular, the {\it normalized colored HOMFLY polynomial} can be represented as the vacuum expectation value of the Wilson loop operator in the 3d Chern-Simons theory:
     \begin{equation}
     \label{WilsonLoopExpValue}
         \mathcal{H}_{R}^{\mathcal{K}}(q, a) = \frac{1}{\text{qdim}(R)}\left\langle \text{tr}_{R} \ \text{Pexp} \left( \oint_{\mathcal{K}} A \right) \right\rangle_{\text{CS}},
     \end{equation}
     where Pexp denotes a path-ordered exponential and the Chern-Simons action is given by
     \begin{equation}\label{CSAction}
         S_{\text{CS}}[A] = \frac{\kappa}{4 \pi} \int_{S^3} \text{tr} \left(  A \wedge dA +  \frac{2}{3} A \wedge A \wedge A \right).
     \end{equation}
     The contour of the Wilson loop operator is a knot $\mathcal{K}$ and $R$ is a representation of the algebra $\mathfrak{sl}_N$, that corresponds to $SU(N)$ gauge group, $\text{qdim}(R)$ is a quantum dimension. In formulas~\eqref{WilsonLoopExpValue} and~\eqref{CSAction} $A=A^kT_k$, where $T_k$ are $\mathfrak{sl}_N$ generators. The answer for \eqref{WilsonLoopExpValue} is a polynomial in two variables $q$ and $a$ that are parametrized as follows:
         $q = e^{\hbar}, \ a = e^{N \hbar}, \ \hbar := \frac{2 \pi i}{\kappa + N}\ $.
     
     One obtains a perturbative expansion in the parameter $\hbar$ of \eqref{WilsonLoopExpValue}:
     \begin{align}
     \begin{aligned}
     \label{LoopExpansionHOMFLY}
      \mathcal{H}_{R}^{\mathcal{K}} = \sum\limits_{n=0}^{\infty} \oint dx_{1}\int
      dx_{2}...\int dx_{n} \langle\,A^{a_1}(x_{1})A^{a_2}(x_{2})...A^{a_n}(x_{n})\,\rangle\,
     \tr(T^{a_1} T^{a_2}...T^{a_n}) \, \hbar^n = \sum_{n=0}^{\infty} \left( \sum_{m=1}^{\text{dim}\,\mathbb{G}_{n}} \mathcal{V}^{\mathcal{K}}_{n,m} \ \mathcal{G}_{n, m}^{R} \right) \hbar^{n},
     \end{aligned}
     \end{align}
     where $\text{dim}\,\mathbb{G}_{n}$ is a number of linearly independent group factors $\mathcal{G}_{n, m}^{R}$ at a certain level $n$. A gauge fixing procedure can be done in different ways (see \cite{Labastida1997uw, GUADAGNINI1990575, Guadagnini:1989kr}), but the resulting polynomial does not depend on it. The main property of the perturbative expansion~\eqref{LoopExpansionHOMFLY} is that knot and representation dependences split. The knot dependent functions $\mathcal{V}^{\mathcal{K}}_{n,m}$ are celebrated \textit{Vassiliev invariants} \cite{chmutov_duzhin_mostovoy_2012}.  The representation dependent functions $\mathcal{G}_{n, m}^{R}$ are called \textit{group factors} (for a precise mathematical definition in Ch.4 and Ch.6 in~\cite{chmutov_duzhin_mostovoy_2012}). Making a perturbation expansion, it can be easily seen that they are the traces of $\mathfrak{sl}_N$ generators $T_k$ in the representation $R$. To provide an example we present explicitly the simplest group factor:
     \begin{equation}
     \label{G21}
         \mathcal{G}_{2, 1}^{R} = \text{tr}_R \left(\sum_{a,b}   T_a T_b T_a T_b  -  T_a T_a T_b T_b \right) = \sum_{a,b,c} f_{abc} \ \text{tr}_R \left( T_a T_c T_b \right) = N \mathcal{C}_2^{R}\,,
     \end{equation}
     where the generators are normalized as $\text{tr}_R \left( T_a T_b \right) = \frac{\delta_{ab}}{2 \dim R} \,$, and $\mathcal{C}_2^{R}$ is an eigenvalue of the quadratic Casimir operator of $\mathfrak{sl}_N$. 
     
     Note that expressions under traces in group factors lie in the center of the universal enveloping algebra $ZU(\mathfrak{sl}_N)$ (for a proof see Ch.6.1.2 in~\cite{chmutov_duzhin_mostovoy_2012}). Casimir operators $\hat{\mathcal{C}}_k, \ k = 1, \ldots, N,$ multiplicatively generate basis elements in $ZU(\mathfrak{sl}_N)$. Hence being the traces and taken in a certain representation $R$, group factors are polynomials of their eigenvalues in the representation $R\,$:
     \begin{equation}\label{GPol}
         \mathcal{G}_{n, m}^{R} =\mathcal{G}_{n,m}\left( \mathcal{C}_1^R, \ldots, \mathcal{C}_{N}^R \right)= \text{Pol}\left( \mathcal{C}_1^R, \ldots, \mathcal{C}_{N}^R \right).
     \end{equation}
     \textbf{Our goal} is \textit{to provide an algorithm} for finding these functions explicitly for all $n,m$ and for any representation $R$.
     
     Multiplication of group factors respects the level, that is the sum of the factors' levels is a level of the resulting group factor. 
     For example, we can set the first group factor of the 4-th level to be the square of the group factor of the second level:
     \begin{equation}
     \mathcal{G}^R_{4,1}=\left(\mathcal{G}^R_{2,1}\right)^2.    
     \end{equation}
      Group factors that cannot be represented as products of other group factors are called \textit{primary group factors}. And Vassiliev invariants corresponding to primary group factors are called \textit{primary} or \textit{primitive Vassiliev invariants}. 
      
           As it has been mentioned in Introduction, there are strong connections and common structures between Wilson lines in $\mathcal{N} = 4$ super Yang-Mills and 3d Chern-Simons theory with $SU(N)$ gauge group. We provide one of basic results of \cite{SYMgroupfactors} to make the connection clear.
     
     In $\mathcal{N} = 4$ SYM the vacuum expectation value of the Wilson line operator $W_R$ in a representation $R$ can be reduced to an average $\frac{1}{\dim R}\langle \tr_R e^{2\pi M} \rangle$ in a Gaussian matrix model. This average has the following perturbative expansion:
\begin{equation}
    \langle W_{R} \rangle = \frac{1}{\dim R} \sum_{n = 0}^{\infty} \frac{(2\pi)^{2n}}{(2n)!}  \langle m_{a_1} m_{a_2} \ldots m_{a_{2n}} \rangle \, \text{tr}_{R} \left( T_{a_1} T_{a_2} \ldots T_{a_{2n}} \right), \hspace{15mm} \langle m_{a} m_{b} \rangle = \frac{g_{YM}^2}{8 \pi^2} \delta_{ab}\,,
\end{equation}
where expressions with traces are group factors. Compare it with~\eqref{LoopExpansionHOMFLY}. This perturbative answer for $\langle W_R \rangle$ is exact due to the absence of the instanton corrections in $\mathcal{N} = 4$ SYM \cite{Pestun}. The HOMFLY group factors turn out to be exactly the same as ones that arise in $\mathcal{N} = 4$ SYM theory.
      

\setcounter{equation}{0}
\section{Explicit construction of group factors}\label{WeightSystem}
As it has been discussed in the previous section group factors are functions of the eigenvalues of $\mathfrak{sl}_N$ Casimir operators. However, a particular form of these functions is currently unknown. We offer an approach to solve this problem, i.e. find explicitly the functions $\mathcal{G}_{n,m}\left( \mathcal{C}_1^R, \ldots, \mathcal{C}_{N}^R \right)$~\eqref{GPol}. The approach includes two ideas. \\
$\bullet$ The first idea is to use an analytical continuation of $\mathfrak{sl}_N$ Casimir eigenvalues. In other words, we utilize the embedding $ZU(\mathfrak{sl}_N)\subset ZU(\mathfrak{gl}_\infty)\,$. We argue that this continuation is the same as the analytical continuation of the quantum $\mathfrak{sl}_N$ invariants to the colored HOMFLY polynomials. \\
$\bullet$ The second idea is to use known symmetries of the colored HOMFLY polynomials to restrict the form of the functions $\mathcal{G}_{n,m}\left( \mathcal{C}_1^R, \ldots, \mathcal{C}_{N}^R \right)$.
\subsection{Analytical continuation of $\mathfrak{sl}_N$ Casimir eigenvalues}
\label{embed into gl}
In this subsection we provide an analytical continuation by rewriting formulas for eigenvalues of $\mathfrak{sl}_N$ Casimir operators in terms of $\mathfrak{gl}_{\infty}$ Casimirs $C_k^{R}$. The latter ones are shifted symmetric functions and defined as follows~\cite{nla.cat-vn1878494}:
\begin{equation}
    \label{gl Casimirs}
    C_{k}^{R} = \sum_{i = 1}^{\infty} \left( R_i - i + 1/2 \right)^k - \left(-i + 1/2 \right)^k.
\end{equation}
Note that the sum is finite for any finite Young diagram $R$, since we set for convenience $R_i = 0$ for sufficiently large $i$. The basis of the Casimir eigenvalues $C_k^R$~\eqref{gl Casimirs} is distinguished by the following facts. The corresponding Hurwitz partition function \cite{Mironov2010yg} becomes a KP $\tau$-function  \cite{Kharchev1993xc,ORLOV200151} and in terms of the Hurwitz partition function, this basis corresponds to the completed cycles and establishes a correspondence with the Gromov-Witten theory \cite{Okounkov2002cja}. 

$\mathfrak{sl}_N$ Casimir eigenvalues $\widetilde{\mathcal{C}}_{k}^{R}$ as functions of $R$ can be obtained from the well-known generating function \cite{Perelomov1968CASIMIROF}:
\begin{equation}\label{gen func sl}
    G_{\mathfrak{sl}_N}(z) = z^{-1} \left( 1 - \prod_{i=1}^{N} \left( 1 - \frac{z}{1 - \lambda_{i} z} \right) \right) = \sum_{k = 0}^{\infty} \widetilde{\mathcal{C}}_{k}^{R} \, z^k, 
\end{equation}
where $\lambda_i = R_i - \frac{1}{N}\sum_{j=1}^N R_j - i + N\,$. $\mathfrak{sl}_N$ Casimir eigenvalues can be expressed in terms of the functions $C_k^R$~\eqref{gl Casimirs}. In what follows we omit $R$ superscripts where it does not lead to misunderstanding. We provide formulas of this kind for the Casimir eigenvalues $\widetilde{\mathcal{C}}^{R}_k$ that are obtained from the generating function~\eqref{gen func sl} in low orders:
\begin{align}\label{CExample}
    \begin{aligned}
        \widetilde{\mathcal{C}}^{R}_2 &= C_2 - \frac{1}{N} C_1^2 + N C_1\,, \\
        \widetilde{\mathcal{C}}^{R}_3 &= C_3 - \frac{3}{N} C_2 C_1 + \frac{2}{N^2} C_1^3 +  2 N C_2 - \frac{7}{2} C_1^2 +\frac{4N^2+1}{4} C_1\, . 
    \end{aligned}
\end{align}
A main feature of these formulas is that they provide an {\it analytic continuation}. Irreducible finite dimensional representations of $\mathfrak{sl}_N$ are enumerated by Young diagrams $R$ with at most $N$ rows, i.e. $l(R) \leqslant N$. Remarkably, formulas for the Casimir eigenvalues in terms of the functions $C_k$ are applicable for any Young diagrams $R$ and values of $N$, while in the sector $l(R) \leqslant N$ their values coincide with the usual $\mathfrak{sl}_N$ Casimir eigenvalues. From theorems of complex analysis and usual arguments concerning an analytical continuation of polynomial functions, the analytical continuation is unique. Hence, the above discussed analytical continuation is the same as the $a = q^N$ continuation that we have discussed at the beginning of Section \ref{colored HOMFLY}.

At the end of this section we discuss translational properties of $\mathfrak{sl}_N$ Casimir eigenvalues which are essential in Subsection~\ref{TTHProof}. Note that the generating function~\eqref{gen func sl} is invariant under {\it translations}: $R_i \rightarrow R_i + \delta R$. This peculiarity corresponds to the following fact from $\mathfrak{sl}_N$ representation theory: Young diagrams $[R_1 + \delta R, R_2 + \delta R, \ldots, R_N + \delta R]$ and $[R_1, R_2, \ldots, R_N]$ correspond to the same irreducible representation. Being translation invariant functions, $\mathfrak{sl}_N$ Casimir eigenvalues $\mathcal{C}_k \left( C^R_1, \ldots, C^R_k \right)$ are invariant under the replacing $C_k^R$ with
\begin{equation}
\label{transl C_k}
\boxed{
    C_{k}^{R + \delta R} = \sum_{j = 0}^{k} (\delta R)^{j} \binom{k}{j} \left(C_{k-j}^R + \theta_{k-j}^{N}\right) - \theta_{k}^N\,,
}
\end{equation}
where we introduce the function $\theta_k^{N}=\sum\limits_{i=1}^N\left(-i+\frac{1}{2}\right)^{k}$ to simplify formulas. 
\subsection{Symmetries of the colored HOMFLY polynomials}\label{HOMFLYsym}
In this subsection we review known symmetries and properties of the colored HOMFLY polynomials. In the following sections we use them to restrict the form of the functions
$\mathcal{G}_{n,m}\left( C_1, \ldots, C_{N} \right)$~\eqref{GPol}.
Group factors depend on $\mathfrak{gl}_{\infty}$ Casimir invariants $C_k$ because $\mathfrak{sl}_N$ Casimir invariants are expressed through them. In particular, we discuss action of the symmetries on the HOMFLY group factors and on the functions $C_k$.\\
$\bullet$ \textbf{Genus order}. The genus order restriction comes from the fact that the genus expansion of the colored HOMFLY polynomials is well defined \cite{Mironov2013oma,Mironov_2013}. The genus order $g$ is defined for a group factor's component as follows: $g\left( N^k C_{\Delta} \right) = k + |\Delta|\,$.
The genus order of a group factor is defined to be the maximal genus order of its components. For example, one can compute the genus order of the simplest group factor:
\begin{equation}
    g \left( \mathcal{G}_{2, 1}^{R} \right) = g \left( N C_2 - C_1^2 + N^2 C_1 \right) = 3\,.
\end{equation}
The genus order of a {\it primary} group factor $\mathcal{G}_{n, m}^{R}$ is bounded: \begin{equation}
\boxed{
     g \left( \mathcal{G}_{n, m}^{R} \right) \leqslant n + 1\,.
    }
\end{equation}
$\bullet$ \textbf{Rank-level duality}. The rank-level duality of the Chern-Simons theory \cite{Naculich1990hg,Naculich1990pa,Mlawer1990uv,Liu2010zs} provides the following relation:
\begin{equation}
\mathcal{H}^\mathcal{K}_R(q,a)=\mathcal{H}^\mathcal{K}_{R^T}(q^{-1},a)\,,
\end{equation}
where $R^T$ is a representation obtained by transposing the Young diagram. This property imposes a condition on group factors: 
\begin{equation}
\label{rank-level-duality}
\boxed{
    \mathcal{G}_{n,m}^{R^T} = (-1)^n \, \mathcal{G}_{n,m}^{R}\Big|_{N\rightarrow-N}\,.
    }
\end{equation}
The Casimir eigenvalues of $\mathfrak{gl}_{\infty}$ transform in a simple way under the transposition of the diagram:
\begin{equation}
\label{Ck(RT)}
    C_k^{R^{T}} = (-1)^{k + 1} C_k^{R}\,.
\end{equation}
$\bullet$ \textbf{Conjugation symmetry}. The conjugation symmetry is defined at a fixed value of $N$. It comes from the representation theory, where a conjugate representation $\overline{R}$ is defined as a complement of a Young diagram $R$ to the rectangular $R_1 \times N$. 
The fact, that the colored HOMFLY polynomials for a representation $R$ and its conjugate $\overline{R}$ coincide, imposes a condition on the group factors:
\begin{equation}
\label{conj sym for group factors}
\boxed{
    \mathcal{G}_{n,m}^{\overline{R}} = \mathcal{G}_{n,m}^{R}\,.
    }
\end{equation}
As a direct corollary of the conjugate transformation, $\overline{R}_{i}=R_{1}-R_{N-i+1}\,$, and the definition of functions $C_k^{R}$ \eqref{gl Casimirs} one can derive the following transformation rule:
\begin{equation}
\begin{aligned}\label{adjC}
    C_{k}^{\overline{R}}=(-1)^k\sum_{p=0}^{k} \ \epsilon^{p} \ \binom{k}{p}\left(C_{k-p}^R+\theta_{k-p}^N \right)-\theta_k^N.
\end{aligned}
\end{equation}
The parameter $\epsilon=N-R_1$ is arbitrary, since formula~\eqref{adjC} is valid for any representation $R$ and value of $N$. We emphasise the presence of the factor $(-1)^{k}$ in front of the sum. The transformation rule is a combination of a translation by $R_1$ and a change $R_i \rightarrow -R_i$.
\\
$\bullet$ \textbf{Tug-the-hook symmetry}. In this section we review the recently discovered tug-the-hook symmetry of the colored HOMFLY polynomials. The tug-the-hook symmetry \cite{mishnyakov2021new, mishnyakov2020novel} of the colored HOMFLY polynomials reads:
\begin{equation}
\label{tth sym}
\mathcal{H}^{\mathcal{K}}_{R}\left(q, a = q^N\right)=\mathcal{H}^{\mathcal{K}}_{\mathbf{T}_{\epsilon}^{N}(R)}\left(q, a = q^N\right),
\end{equation}
where $\mathbf{T}_{\epsilon}^{N}$ is a transformation of a Young diagram which pulls the diagram inside the $(N+M|M)$ fat hook.
This symmetry has the supergroup origin \cite{mishnyakov2020novel}. The Reshetikhin-Turaev quantum knot invariants for quantum superqroup $U_q(\mathfrak{sl}(N|M))$ and quantum group $U_q(\mathfrak{sl}(|N-M|)$ exactly coincide. The tug-the-hook symmetry is a reincarnation of the equivalence relation in the representation theory of the supergroup $\mathfrak{sl}(N|M)$.
To describe the symmetry quantitatively one can use an analogue of Frobenius notation for Young diagrams:
\begin{equation}\label{TugTheHookYoungTNot}
\left[R_{1}, \ldots, R_{N}\right]\left(\alpha_{N+1}, \ldots, \alpha_{N+M} \mid \beta_{N+1}, \ldots, \beta_{N+M}\right),
\end{equation}
where $R_{i}$, $1 \leq i \leq N$, are lengths of the first $N$ rows of the Young diagram $R$ and the rest of the diagram is parametrized by shifted Frobenius variables:
\begin{equation}
\alpha_{i}=R_{i}-(i-N)+1, \quad
\beta_{i}=R_{i-N}^{T}-i+1, \quad i>N.
\end{equation}
Then $\mathbf{T}_{\epsilon}^{N}$ is the following transformation: $R_{i} \rightarrow R_{i}+\epsilon, \, \alpha_{i} \rightarrow \alpha_{i}+\epsilon, \, \beta_{i} \rightarrow \beta_{i}-\epsilon,$ where $\epsilon$ is an integer, such that the result is still a Young diagram. 
One can rewrite formulas for Casimir invariants \eqref{gl Casimirs} in notation \eqref{TugTheHookYoungTNot}:
\begin{equation}\label{TTHC}
C_{k}^{R}=\sum_{i=1}^N \left(R_{i}-i+1/2\right)^{k}-\left(-i+1/2\right)^{k} +\sum\limits_{i=N+1}^{N+M} \left(\alpha_i-N-1/2\right)^k-\left(-\beta_i-N+1/2\right)^k .
\end{equation}
Using this expression one derives transformation rules for functions $C_k^R$ under the action of the tug-the-hook symmetry: 
\begin{equation}\label{tth transf}
\boxed{
C_{k}^{\mathbf{T}^N_{\epsilon}(R)}=\sum_{p=0}^{k} \epsilon^{p} \binom{k}{p} \left(C_{k-p}^{R}+\theta_{k-p}^N\right)-\theta_{k}^N\,.
}
\end{equation}

\subsection{HOMFLY group factors}\label{ConstrWeightSystem}
In the previous subsection we have preliminary discussed the embedding $Z U(\mathfrak{sl}_N)\subset Z U(\mathfrak{gl}_\infty)$ and restrictions imposed on the HOMFLY group structure by the known symmetries of the colored HOMFLY polynomials.

An explicit description of the embedding $Z U(\mathfrak{sl}_N)\subset Z U(\mathfrak{gl}_\infty)$ can be provided by the generating function~\eqref{gen func sl}. However, it is more convenient to describe this embedding with the use of an analytical formula for the $n$-th term of the sum:
\begin{equation}\label{AnalyticZ}
    \boxed{\mathcal{C}^R_{n}=\sum\limits_{m=0}^{n}\frac{(-1)^{n-m}}{N^m}\binom{n}{m}  
    \left(
    \left(C^R_{n-m}+\theta_{n-m}^N \right) \left(C^R_1+\theta_1^N \right)^m-\theta_{n-m}^N\left(\theta_1^N\right)^m 
    \right)\,,}
\end{equation}
which can be obtained from the translation invariance of $\mathfrak{sl}_N$ representations. Remind that $\theta_k^{N}=\sum\limits_{i=1}^N\left(-i+\frac{1}{2}\right)^{k}$. Note that the Casimir invariants $\mathcal{C}_{n}$ are linear combinations of the $\widetilde{\mathcal{C}}_k$, that comes from the generating function \eqref{gen func sl}. For example, one can easily see by a direct comparison with~\eqref{CExample} that $\mathcal{C}_{2}=\widetilde{\mathcal{C}}_2$ and $\mathcal{C}_{3}=-\widetilde{\mathcal{C}}_3+\frac{N}{2}\widetilde{\mathcal{C}}_2$. 

Group factors of the colored HOMFLY polynomials must be well-defined for their special case -- the colored Alexander polynomials, i.e. group factors must not have singularity at $N\rightarrow 0\,$. We choose a special basis in the center of the universal enveloping algebra $ZU(\mathfrak{sl}_N)$ that respects the symmetries of the colored HOMFLY polynomials, as it has been described in Subsection~\ref{HOMFLYsym}. Namely, we present two types of Casimir invaraints that appear in the perturbative expansion, the so-called even and odd elements.\\
$\bullet$ The \textit{even elements} are denoted as $\mathcal{C}^R_{[2n]}$ and have the following form:
    \begin{equation}\label{AnalyticZeven}
    \mathcal{C}^R_{[2n]}:=\sum\limits_{k=1}^{2n}(-1)^k\frac{C^R_k \left(C^R_{2n-k}+2\theta^N_{2n-k}\right)}{2\cdot k!(2n-k)!}\,.
\end{equation}
$\bullet$ We provide an algorithm to construct the \textit{odd elements}. Note that $\mathcal{C}^R_{2n+1}$  change their signs with respect to the conjugation symmetry, so in group factors they are present only in even combinations, $\mathcal{C}^R_{2 n_1 + 1}\mathcal{C}^R_{2 n_2 + 1} $.  
We get rid of negative powers of $N$ from formulas for $\mathcal{C}^R_{2n_1+1}\mathcal{C}^R_{2n_2+1}$ by multiplying them on appropriate powers of $N$ and then adding linear combinations of Casimir invariants. This algorithm is done for each case separately. We denote the resulting \textit{not singular} in $N$ expressions as $\mathcal{C}^R_{[2n_1+1,2n_2+1]}$. Provide as an example the formulas for $\mathcal{C}^R_{[3,3]}$ and $\mathcal{C}^R_{[5,3]}$: 
\begin{equation}\label{ZOdd}
{\small
\begin{aligned}
\mathcal{C}^R_{[3,3]}&:=\frac{1}{N}\left(\frac{8\,\mathcal{C}_{[2]}^3+\frac{1}{4}N^4\mathcal{C}_{3}^2}{N^2}-\mathcal{C}_{[2]}^2-12\,\mathcal{C}_{[2]} \mathcal{C}_{[4]}\right), \\
\mathcal{C}^R_{[5,3]}&:=\frac{1}{N^3}\left(\frac{\frac{1}{16} \mathcal{C}_{5} \mathcal{C}_{3} N^6-8\, \mathcal{C}_{[2]}^4}{N^2}+\frac{4}{3} \mathcal{C}_{[2]}^3+36\, \mathcal{C}_{[4]} \mathcal{C}_{[2]}^2\right)-\frac{2\, \mathcal{C}_{[2]}\mathcal{C}_{[3,3]}}{N^2}
-\frac{1}{N}\left(\frac{10}{3} \mathcal{C}_{[2]}^3-\frac{\mathcal{C}_{[2]}^2}{24}-4\, \mathcal{C}_{[4]} \mathcal{C}_{[2]}-30\, \mathcal{C}_{[6]} \mathcal{C}_{[2]}-24\, \mathcal{C}_{[4]}^2\right).
\end{aligned}
}
\end{equation}
The two types of functions, $\mathcal{C}^R_{[2n]}$ and $\mathcal{C}^R_{[2n_1+1,2n_2+1]}$, multiplicatively generate Casimir elements that are included in the HOMFLY group structure. We denote these Casimir elements by $\mathcal{C}_{\Delta}$, where $\Delta$ is a Young diagram without unit entries. For example, $\mathcal{C}_{[3,3,2]} = \mathcal{C}_{[3,3]} \mathcal{C}_{[2]}$.
So, in what follows we describe the HOMFLY group structure in the basis of $\mathcal{C}_\Delta$:
\begin{equation}\label{HLoopExp}
    \mathcal{H}^{\mathcal{K}}_{R}=\sum\limits_{n=0}^{\infty}\hbar^n\sum\limits_{|\Delta|\leq n}\mathcal{C}_{\Delta}^{R}\sum_{m=0}^{n-|\Delta|} \left(v^{\mathcal{K}}_{\Delta,m}\right)_n N^m\,,
\end{equation}
where $\left(v^{\mathcal{K}}_{\Delta,m}\right)_n$ are also Vassiliev invariants and are linear combinations of $\mathcal{V}_{n,m}$ in~\eqref{LoopExpansionHOMFLY}.

In Subsection~\ref{HOMFLYsym} we have found how each symmetry transforms the HOMFLY group factors and the $\mathfrak{gl}_\infty$ Casimir invariants. Knowing these facts, we can now state how the HOMFLY symmetries descends to expansion~\eqref{HLoopExp} and emphasize crucial properties of the basis polynomials~\eqref{AnalyticZeven},~\eqref{ZOdd} with respect to these symmetries.
\begin{enumerate}
        \item The tug-the-hook symmetry does not impose restrictions on~\eqref{HLoopExp}, as it manifests in the same way as the translation invariance of the $\mathfrak{sl}_N$ representations (compare~\eqref{transl C_k} and~\eqref{tth transf}).
    \item The conjugation symmetry is crucial. It forbids $\mathcal{C}_\Delta$ with odd size of Young diagram $|\Delta|$ in~\eqref{HLoopExp} due to the presence of the multiplier $(-1)^k$ in~\eqref{adjC}.
    \item The rank-level duality makes some of the coefficients $\left(v^{\mathcal{K}}_{\Delta,m}\right)_n$ to be zero  due to~\eqref{Ck(RT)}. In fact, the $\mathcal{C}_{[2n]}$ are even with respect to the rank-level transformation and the $\mathcal{C}_{[2n_1+1,2n_2+1]}$, on the contrary, change their signs.
    \item The order restriction puts the upper bound $n-|\Delta|$ in the sum over $m$ and can vanish some $\left(v^{\mathcal{K}}_{\Delta,m}\right)_n$ with biggest $|\Delta|$ and $m$ if the genus order of $N^m\mathcal{C}_{\Delta}^{R}$ turns out to be too high. This is caused by the fact that the genus orders $g\left(\mathcal{C}_{[2n]}\right)=2n+1$ and $g\left(\mathcal{C}_{[2n_1+1,2n_2+1]}\right)=2n_1+2n_2+3\,$.
\end{enumerate}
Note that the properties of $\mathcal{C}_{[2n_1+1,2n_2+1]}$ are just computational observations and actually they can break for higher levels.

Now let us state explicitly which \textbf{group structure of the colored HOMFLY polynomials} follows from the described properties of these basis polynomials and from the rules following from the HOMFLY symmetries.
\begin{itemize}
    \item \textbf{At the even level 2n}, there should be polynomials of the form $N^{2l}\mathcal{C}_{\Delta_e}$, where $\Delta_e$'s are partitions of the even number $m=2, \dots, 2n$ into even terms, and $2l+|\Delta_e| \leq 2n$, and also polynomials $N^{2l+1}\mathcal{C}_{\Delta_o}$, where $\Delta_o$'s are partitions of the even number $m = 2, \dots, 2n$, which contain at least two odd numbers and which do not contain $1$, and $2l + |\Delta_o| +1 \leq 2n-1$.
    \item \textbf{At the odd level 2n+1}, there should be polynomials of the form $N^{2l + 1} \mathcal{C}_{\Delta_e}$, $2l + 1 + |\Delta_e| \leq 2n + 1$, and polynomials $N^{2l}\mathcal{C}_ {\Delta_o}$, $2l + |\Delta_o| \leq 2n$.
\end{itemize}
Thus, using the rules above one can write down the group factors, which can be met in the HOMFLY loop expansion, \textit{up to any level} $n$. And it would be a great success if the known HOMFLY symmetries were constraining enough to fully fix the HOMFLY group structure. So, let us check this conjecture. Namely, fix a knot $\mathcal{K}$ and solve the system of linear equations on each level $n$ with unknown $\left(v^{\mathcal{K}}_{\Delta,m}\right)_n$:
\begin{equation}\label{HExp}
    \sum\limits_{|\Delta|\leq n}\mathcal{C}_{\Delta}^{R_\alpha}\sum_{m=0}^{n-|\Delta|} \left(v^{\mathcal{K}}_{\Delta,m}\right)_n N^m=\left(\mathcal{H}^{\mathcal{K}}_{R_\alpha}\right)_n\,,
\end{equation}
for some set of representations $R_\alpha$. Here $\left(\mathcal{H}^{\mathcal{K}}_{R_\alpha}\right)_n$ is the $\hbar^n$ term in the HOMFLY polynomial expansion $\mathcal{H}^{\mathcal{K}}_{R_\alpha}$. 

Now let us \textit{provide a concrete example} of the described procedure. We have stated above which group factors can be included in the HOMFLY expansion due to the known symmetries restrictions. First, list them up to the 9-th order:
{\small \begin{equation}\label{GroupFactors}
\renewcommand{\arraystretch}{2}
    \begin{tabular}{| m{1em} | m{23em} | m{1em} | m{23em} |} 
        \hline
        $\hbar^2$ & $\mathcal{C}_{[2]}$ & $\hbar^3$ &  $N\mathcal{C}_{[2]}$\\
        \hline 
        $\hbar^4$ & $\cancel{\boxed{\mathcal{C}_{[2]}}}$, $N^2\mathcal{C}_{[2]}$, $\mathcal{C}_{[2]}^2$, $\mathcal{C}_{[4]}$ 
        & $\hbar^5$ & $N\mathcal{C}_{[2]}$, $N^3\mathcal{C}_{[2]}$, $N\mathcal{C}^2_{[2]}$, $N\mathcal{C}_{[4]}$ \\
        \hline
        $\hbar^6$ &$\cancel{\boxed{\mathcal{C}_{[2]}}}$, $N^2\mathcal{C}_{[2]}$, $N^4\mathcal{C}_{[2]}$, $\cancel{\boxed{\mathcal{C}_{[2]}^2}}$, $N^2\mathcal{C}_{[2]}^2$, $\mathcal{C}_{[2]}^2 + 6\mathcal{C}_{[4]}$, $N^2\mathcal{C}_{[4]}$, $\mathcal{C}_{[2]}^3$, $\mathcal{C}_{[2]}\mathcal{C}_{[4]}$, $\mathcal{C}_{[6]}$ 
        &$\hbar^7$ & $N\mathcal{C}_{[2]}$, $N^3\mathcal{C}_{[2]}$, $N^5\mathcal{C}_{[2]}$, $N\mathcal{C}_{[2]}^2$ $N^3\mathcal{C}_{[2]}^2$, $ N\mathcal{C}_{[4]}$, $N^3\mathcal{C}_{[4]}$, $N\mathcal{C}_{[2]}^3$, $N\mathcal{C}_{[2]}\mathcal{C}_{[4]}$, $ \mathcal{C}_{[3,3]}$, $N\mathcal{C}_{[6]}$ \\
        \hline 
        $\hbar^8$ &$\cancel{\boxed{\mathcal{C}_{[2]}}}$, $N^2\mathcal{C}_{[2]}$, $N^4\mathcal{C}_{[2]}$, $N^6\mathcal{C}_{[2]}$, $\cancel{\boxed{\mathcal{C}_{[2]}^2}}$, $N^2\mathcal{C}_{[2]}^2$, $N^4\mathcal{C}_{[2]}^2$, $\mathcal{C}_{[2]}^2+6\mathcal{C}_{[4]}$, $N^2\mathcal{C}_{[4]},\; N^4\mathcal{C}_{[4]}$, $\mathcal{C}_{[2]}^3$, $N^2\mathcal{C}_{[2]}^3$, $\mathcal{C}_{[2]}\mathcal{C}_{[4]}$, $N^2\mathcal{C}_{[2]}\mathcal{C}_{[4]}$, $\mathcal{C}_{[6]}$,
        $N^2\mathcal{C}_{[6]}$, $N\mathcal{C}_{[3,3]}$, $\mathcal{C}_{[2]}^4$ $\mathcal{C}_{[2]}^2\mathcal{C}_{[4]}$, $\mathcal{C}_{[4]}^2$, $\mathcal{C}_{[2]}\mathcal{C}_{[6]}$, $\mathcal{C}_{[8]}$ 
        & $\hbar^9$ & $N\mathcal{C}_{[2]}$, $N^3\mathcal{C}_{[2]}$, $N^5\mathcal{C}_{[2]}$, $N^7\mathcal{C}_{[2]}$, $N\mathcal{C}_{[2]}^2$, $N^3\mathcal{C}_{[2]}^2$, $N^5\mathcal{C}_{[2]}^2$, $N\mathcal{C}_{[4]}$, $N^3\mathcal{C}_{[4]}$, $N^5\mathcal{C}_{[4]}$, $N\mathcal{C}_{[2]}^3$, $N^3\mathcal{C}_{[2]}^3$, $N\mathcal{C}_{[2]}\mathcal{C}_{[4]}$, $N^3\mathcal{C}_{[2]}\mathcal{C}_{[4]}$, $N\mathcal{C}_{[6]}$, $N^3\mathcal{C}_{[6]}$, $\mathcal{C}_{[3,3]}$, $N^2\mathcal{C}_{[3,3]}$, $N\mathcal{C}_{[2]}^4$, $N\mathcal{C}_{[2]}^2\mathcal{C}_{[4]}$, $N\mathcal{C}_{[4]}^2$, $\mathcal{C}_{[2]}\mathcal{C}_{[3,3]}$, $N\mathcal{C}_{[2]}\mathcal{C}_{[6]}$, $N\mathcal{C}_{[8]}$, $\mathcal{C}_{[5,3]}$ \\
        \hline
    \end{tabular}
\renewcommand{\arraystretch}{1}
\end{equation}}
Second, box the group factors which do not appear in the HOMFLY expansions~\eqref{HExp}. This is exactly the result of solving systems~\eqref{HExp} for $n=2,\dots,9$. In other words, the group factors, that are not crossed out, exactly appear in the HOMFLY loop expansion~\eqref{HLoopExp}. Note that one can proceed this way \textit{up to any level} $n$ if knowing enough HOMFLY polynomials $\mathcal{H}^{\mathcal{K}}_{R_\alpha}$ for different representations $R_{\alpha}$. 


\setcounter{equation}{0}
\section{Implications}\label{Implications}
The concrete form of the group factors \eqref{GroupFactors} opens wide perspectives of implications and future research directions. In this section we discuss three consequences which immediately follow from our analysis. 

\subsection{Proof of the tug-the-hook symmetry}\label{TTHProof}
Here we present simple arguments that prove the existence of the recently discovered tug-the-hook symmetry of the colored HOMFLY polynomials \eqref{tth sym}.
\begin{enumerate}
    \item Colored HOMFLY polynomials admit the perturbative expansion \eqref{LoopExpansionHOMFLY}, where the group factors are functions of $\mathfrak{sl}_N$ Casimir eigenvalues $\mathcal{G}_{n,m}^R = \mathcal{G}_{n,m}\left( \mathcal{C}_1^R, \ldots, \mathcal{C}_{N}^R \right)$.
    To show that the colored HOMFLY polynomials have the tug-the-hook symmetry is sufficient to show that group factors are invariant under the action of the symmetry. 
    \item $\mathfrak{sl}_N$ Casimir eigenvalues $\mathcal{C}_k^R$ can be represented as polynomials of $\mathfrak{gl}_{\infty}$ Casimir eigenvalues $C_k^R$ (for details see Section \ref{embed into gl}), that provides the analytic continuation from the sector $l(R) \leqslant N$ to arbitrary Young diagrams. 
    The tug-the-hook symmetry is present only for the colored HOMFLY polynomials as it explicitly involves evaluation $a = q^{N}$ for $N \leqslant l(R)$ \eqref{tth sym}. Therefore the analytic continuation is essential for analysis of the tug-the-hook symmetry. 
    \item The Casimir eigenvalues $\mathcal{C}_k^R$ are translation invariant functions with respect to Young diagrams: $\mathcal{C}_k^{R} = \mathcal{C}_k^{R + \delta R}$. This fact follows directly from the $\mathfrak{sl}_N$ representation theory. 
    \item The $\mathfrak{gl}_{\infty}$ Casimir eigenvalues \eqref{gl Casimirs} transform under the translations $[R_1 , R_2, \ldots, R_N] \rightarrow [R_1 + \delta R, R_2 + \delta R, \ldots, R_N + \delta R]$ by rule \eqref{transl C_k}.
    \item The $\mathfrak{gl}_{\infty}$ Casimir eigenvalues \eqref{gl Casimirs} transform under the tug-the-hook symmetry by rule \eqref{tth transf}.
    \item The translation~\eqref{transl C_k} and the tug-the-hook actions~\eqref{tth transf} on the functions $C_k^R$ coincide for $\delta R = \epsilon$.
    \item From the p.2-p.6 we conclude that the $\mathfrak{sl}_N$ Casimir eigenvalues are invariant under the tug-the-hook transformations: $\mathcal{C}^{R}_k(C_1, \ldots, C_k) = \mathcal{C}^{\mathbf{T}^N_{\epsilon}(R)}_k(C_1, \ldots, C_k)$.
    \item From p.1 and p.7 it follows that group factors are also invariant under the tug-the-hook symmetry since they are functions of the $\mathfrak{sl}_N$ Casimir eigenvalues, in other words,
        $\mathcal{G}_{n,m}^R = \mathcal{G}_{n,m}^{\mathbf{T}^N_{\epsilon}(R)}$. 
        \\
        Hence  {\it the colored HOMFLY polynomial itself has the tug-the-hook symmetry}.
\end{enumerate}

\subsection{Novel conjectural symmetries}\label{NewSymm}
A closer look on the group factors \eqref{GroupFactors} reveals interesting structures, that were not noticed before. Namely, one can note that $\mathcal{C}_{[2]}$ is not included at levels above the second, and $\mathcal{C}_{[2]}^2$ and $\mathcal{C}_{[4]}$ are included only in combination $\mathcal{C}_{[2]}^2 + 6\,\mathcal{C}_{[4]}$ at levels above the fourth. Thus, the known HOMFLY symmetries turn out to be insufficient to fully fix the group structure of the colored HOMFLY polynomials, as there appear the described exceptions from the written in Subsection~\ref{ConstrWeightSystem} rules. Moreover, we assume that there are lots of absent polynomials at higher levels.

This indicates the existence of still unknown hidden symmetries of the HOMFLY polynomials. And finding out these properties is one of goals of our future studies.

\subsection{More Vassiliev invariants}\label{MoreVI}
Possibility of finding Vassiliev invariants was restricted by laboriousness
of calculation of the group factors $\mathcal{G}_{n,m}$ directly from the perturbative computation of the Feynman diagrams. Our proposed method for decomposing the HOMFLY polynomials into the special basis of the $\mathfrak{sl}_N$ Casimir invariants described in Subsection~\ref{ConstrWeightSystem} allows one to proceed further in getting Vassiliev invariants.

We list found by us Vassiliev invariants for the knots $3_1$ (up to the 11-th order) and $5_2$ (up to the 10-th order) on our cite~\cite{knotebook}.
We emphasize that these Vassiliev invariants are written in the basis of group factors different from~\eqref{LoopExpansionHOMFLY}, so that the $v_{n,m}$ are linear combinations of the $\mathcal{V}_{n,m}$, which one can find in~\cite{katlas}. 

\setcounter{equation}{0}
\section{Acknowledgements}\label{Acknowledgements}
We would like to thank A.Yu. Morozov, V.V. Mishnyakov and A.V. Popolitov for useful discussions and interesting questions which motivate us for future research. We are grateful to E. Guadagnini for a very useful correspondence. This work was funded by the Russian Science Foundation (Grant No.20-71-10073).

\printbibliography
\end{document}